\documentclass{article}
\usepackage{amsmath, amssymb, amsfonts}
\usepackage{epsfig}
\title{Boundary Stress Tensors for Spherically - symmetric Conformal Rindler Observers}
\author{Hristu Culetu, \\Ovidius University, Dept.of Physics, \\B-dul Mamaia 124, 900527 Constanta, Romania, \\e-mail : hculetu@yahoo.com}

\begin{document}
\numberwithin{equation}{section}
\pagenumbering{arabic}
\maketitle
\newcommand{\fv}{\boldsymbol{f}}
\newcommand{\tv}{\boldsymbol{t}}
\newcommand{\gv}{\boldsymbol{g}}
\newcommand{\OV}{\boldsymbol{O}}
\newcommand{\wv}{\boldsymbol{w}}
\newcommand{\WV}{\boldsymbol{W}}
\newcommand{\NV}{\boldsymbol{N}}
\newcommand{\hv}{\boldsymbol{h}}
\newcommand{\yv}{\boldsymbol{y}}
\newcommand{\RE}{\textrm{Re}}
\newcommand{\IM}{\textrm{Im}}
\newcommand{\rot}{\textrm{rot}}
\newcommand{\dv}{\boldsymbol{d}}
\newcommand{\grad}{\textrm{grad}}
\newcommand{\Tr}{\textrm{Tr}}
\newcommand{\ua}{\uparrow}
\newcommand{\da}{\downarrow}
\newcommand{\ct}{\textrm{const}}
\newcommand{\xv}{\boldsymbol{x}}
\newcommand{\mv}{\boldsymbol{m}}
\newcommand{\rv}{\boldsymbol{r}}
\newcommand{\kv}{\boldsymbol{k}}
\newcommand{\VE}{\boldsymbol{V}}
\newcommand{\sv}{\boldsymbol{s}}
\newcommand{\RV}{\boldsymbol{R}}
\newcommand{\pv}{\boldsymbol{p}}
\newcommand{\PV}{\boldsymbol{P}}
\newcommand{\EV}{\boldsymbol{E}}
\newcommand{\DV}{\boldsymbol{D}}
\newcommand{\BV}{\boldsymbol{B}}
\newcommand{\HV}{\boldsymbol{H}}
\newcommand{\MV}{\boldsymbol{M}}
\newcommand{\be}{\begin{equation}}
\newcommand{\ee}{\end{equation}}
\newcommand{\ba}{\begin{eqnarray}}
\newcommand{\ea}{\end{eqnarray}}
\newcommand{\bq}{\begin{eqnarray*}}
\newcommand{\eq}{\end{eqnarray*}}
\newcommand{\pa}{\partial}
\newcommand{\f}{\frac}
\newcommand{\FV}{\boldsymbol{F}}
\newcommand{\ve}{\boldsymbol{v}}
\newcommand{\AV}{\boldsymbol{A}}
\newcommand{\jv}{\boldsymbol{j}}
\newcommand{\LV}{\boldsymbol{L}}
\newcommand{\SV}{\boldsymbol{S}}
\newcommand{\av}{\boldsymbol{a}}
\newcommand{\qv}{\boldsymbol{q}}
\newcommand{\QV}{\boldsymbol{Q}}
\newcommand{\ev}{\boldsymbol{e}}
\newcommand{\uv}{\boldsymbol{u}}
\newcommand{\KV}{\boldsymbol{K}}
\newcommand{\ro}{\boldsymbol{\rho}}
\newcommand{\si}{\boldsymbol{\sigma}}
\newcommand{\thv}{\boldsymbol{\theta}}
\newcommand{\bv}{\boldsymbol{b}}
\newcommand{\JV}{\boldsymbol{J}}
\newcommand{\nv}{\boldsymbol{n}}
\newcommand{\lv}{\boldsymbol{l}}
\newcommand{\om}{\boldsymbol{\omega}}
\newcommand{\Om}{\boldsymbol{\Omega}}
\newcommand{\Piv}{\boldsymbol{\Pi}}
\newcommand{\UV}{\boldsymbol{U}}
\newcommand{\iv}{\boldsymbol{i}}
\newcommand{\nuv}{\boldsymbol{\nu}}
\newcommand{\muv}{\boldsymbol{\mu}}
\newcommand{\lm}{\boldsymbol{\lambda}}
\newcommand{\Lm}{\boldsymbol{\Lambda}}
\newcommand{\opsi}{\overline{\psi}}
\renewcommand{\tan}{\textrm{tg}}
\renewcommand{\cot}{\textrm{ctg}}
\renewcommand{\sinh}{\textrm{sh}}
\renewcommand{\cosh}{\textrm{ch}}
\renewcommand{\tanh}{\textrm{th}}
\renewcommand{\coth}{\textrm{cth}}

\begin{abstract}
 The boundary energy - momentum tensors for a static observer in the conformally flat Rindler geometry are considered. We find that the surface energy density is positive far from the Planck world, but that the transversal pressures are negative. The kinematical parameters associated with the nongeodesic congruence of static observers are computed. The entropy $S$ corresponding to the degrees of freedom on the 2 - surface of constant $\rho$ and $t$ equals the horizon entropy of a black hole with a time - dependent mass, and the Padmanabhan expression $E = 2 S T$ is obeyed. The 2 -  surface shear tensor is vanishing, and the coefficient of the bulk viscosity $\zeta$ is $1/16 \pi$, so the negative pressure due to it acts as a surface tension.

\textbf{Pacs} : 04.50.+h, 04.70.Dy, 04.90.+e, 05.90.+m

\textbf{Keywords}: Surface energy, Conformal Rindler geometry, Horizon entropy, Equipartition law
\end{abstract}

\maketitle

\section{INTRODUCTION}

 There has been renewed interest in recent years in boundary matter. General relativity, like Newtonian gravitation, indicates
  that matter is the source for gravity. 
 Khoury and Parikh \cite{KP} asked the question whether the acceleration could be attributed to matter: where are energy and stresses localized as we could accelerate even in Minkowski space? We know that inertial frames exist even in the total absence of matter (the matter distribution is encoded in the stress tensor). Usually the boundary conditions for the metric are in the form of an induced metric and an extrinsic curvature for some hypersurface. 
 
 The boundary matter concept appears, for example, in the so-called membrane paradigm for a black - hole's stretched horizon \cite{TPM,PW,HC1} viewed as an inner boundary. It is worth noting that the Gibbons - Hawking term in the Einstein - Hilbert action is a surface integral over the outer boundary of spacetime.
 
 As Khoury and Parikh have conjectured, ''matter refers to both bulk and boundary stress tensors.'' They uniquely specify the geometry of spacetime. This may be recognized from the fact that the bulk stress tensor does not fully determine the Riemann tensor which, besides the Ricci tensor, contains the Weyl tensor (its free of a trace part). In addition, the Weyl tensor is independent of matter. Therefore, the authors of Ref. 1 emphasized that, to determine the solutions of Einstein's equations, one needs to add boundary conditions. Other contexts where boundary matter arises are the brane-world scenarios, AdS - CFT correspondence \cite{OA}, and the Brown - York construction of a gravitational boundary stress tensor \cite{BY}. 
 
 The purpose of this paper is to compute the boundary stress tensors for a uniformly accelerating distribution of conformal Rindler observers \cite{HC2, HC3}. Our ''boundary sources'' are located on any $\rho = const.$ hypersurface, where $\rho$ is the radial coordinate. As Padmanabhan \cite{TP} has observed, one can associate to the area $A$ of a spherical surface around a massive body the degrees of freedom corresponding to the entropy $S = A/4$ because observers at rest on the surface will have an acceleration produced by the body. In addition, for accelerated observers, one can, in principle, locate holographic screens (equipotential surfaces) anywhere in space \cite{EV}.
 
In Section II, we introduce the conformally - flat Rindler metric and its connection with Minkowski's spacetime. We further depict the Penrose diagram of the spacetime and its causal structure. Section III treats the kinematical parameters associated with a nongeodesic congruence of static observers in the same geometry, paying special attention to the observer's acceleration and its direct relation to the energy density located on a $\rho = $ const. hypersurface. In addition, an Unruh temperature is introduced, keeping in mind that a ''static'' observer is in fact accelerating. Section IV is the main section of the paper. On the grounds of the Parikh and Wilczek's \cite{PW} and Eling and Oz \cite{EO} papers, we compute the energy momentum tensor on a $\rho = \rho_{0}$ timelike hypersurface and, from there, the surface energy density $\Sigma(\rho)$. We found that $\Sigma$ undergoes fluctuations near the Planck world, changing even its sign at $\rho = l_{P}$. A similar sign change takes place for the energy enclosed by a two surface $H$ of constant $\rho$ and $t$. In addition, $H$ is endowed with a positive - bulk - viscosity and negative -  pressure Newtonian fluid.

The new effects found in this paper refer to the behavior of the acceleration $a(\rho)$ of a static observer, the proper temperature associated to it, the surface energy density $\Sigma$ on a $\rho = \rho_{0}$ hypersurface and the energy $E$ enclosed by the 2 - surface $\rho = \rho_{0}, t = t_{0}$. All of them vanish at $\rho = b$ (the Planck length). Moreover, $a^{\rho}, \Sigma$ and $E$ experience a sign change for that value of $\rho$. Throughout the paper we use the geometrical units $G = c = \hbar = 1$ .

\section{CONFORMALLY - FLAT RINDLER METRIC}

We introduce in this section the conformal Rindler observer in spherical coordinates, emphasizing the special role played by the conformal factor. As we shall see, far from the Planck world ($\rho >> b$), the geometry is an ordinary Rindler space in spherical coordinates. Therefore, it is worth studying the boundary stress tensors in this geometry and seeing what new effects arise due to the presence of the conformal factor.
  
 Let us consider the 4 - dimensional subspace of the Witten bubble spacetime \cite{EW}
 \begin{equation}
  ds^{2} = -g^{2} r^{2} dt^{2} + (1-\frac{4 b^{2}}{r^{2}})^{-1} d r^{2} + r^{2} \text{cosh}^{2} gt~ d\Omega^{2} 
  \label{1}
  \end{equation}
  with $b$ and $g$ being constants and $d\Omega^{2} = d\theta^{2} + \text{sin}^{2}\theta d\phi^{2}$ being the metric on the unit 2 - sphere. The metric in Eq. (1) is the ordinary Minkowski space provided $r>>2b$, but written in spherical Rindler coordinates \cite{HC4}. The spherical distribution of uniformly accelerated observers with the rest - system acceleration $g$ uses this type of hyperbolically expanding coordinates. The singularity at $r = 2b$ is only a coordinate singularity, as can be seen from the isotropical form of Eq. (1) with the help of a new radial coordinate $\rho$
 \begin{equation}
 r = \rho + \frac{b^{2}}{\rho}.
 \label {2}
 \end{equation}
 Therefore, the spacetime in Eq. (1) now appears as 
  \begin{equation}
   ds^{2} = \left( 1+\frac{b^{2}}{\rho^{2}}\right)^{2} (-g^{2} \rho^{2} dt^{2} + d\rho^{2} + \rho^{2} \text{cosh}^{2} gt~ d\Omega^{2}).
 \label{3}
 \end{equation}
 As Cho and Pak \cite{CP} have observed, the singularity at $\rho = 0$ is a coordinate singularity because the curvature invariant $R_{\alpha \beta}R^{\alpha \beta}$ and the Kretschmann scalar $R_{\alpha \beta \mu \nu}R^{\alpha \beta \mu \nu}$ are finite there. Therefore, the conformally flat metric in Eq. (3) can be interpreted as an instanton - antiinstanton pair in conformal gravity \cite{CP}.
 
  We note that the coordinate transformation 
  \begin{equation}
 \bar{x} = \rho ~\text{cosh} gt~ \text{sin}\theta ~\text{cos} \phi ,~~\bar{y} = \rho ~\text{cosh} gt~ \text{sin} \theta~ \text{sin} \phi,~~\bar{z} = \rho~ \text{cosh} gt ~\text{cos}\theta, ~~\bar{t} = \rho ~\text{sinh} gt
 \label{4}
 \end{equation}
changes the previous metric to
\begin{equation}
ds^{2} = ( 1+\frac{b^{2}}{\bar{x}_{\alpha} \bar{x}^{\alpha}})^{2} ~\eta_{\mu \nu}~ d\bar{x}^{\mu}~ d\bar{x}^{\nu},
\label{5}
\end{equation}
which is conformally flat ( $\eta_{\mu \nu} = diag (-1, +1, +1, +1)$ and the Greek indices run from $0$ to $3$), and $\rho^{2} = \bar{x}^{i} \bar{x}_{i} - \bar{t}^{2} \equiv \bar{R}^{2} - \bar{t}^{2} , i = 1,2,3$. From now on we take the constant $b$ to be of the order of the Planck length.

Let us observe that the geometry in Eq. (3) becomes flat provided $\rho>>b$ (the conformal factor tends to unity ) or $\rho<<b$ when the first term in the conformal factor may be neglected. Therefore, Eq. (3) represents the Lorentzian version of the Euclidean Hawking wormhole \cite{SWH} \cite{SW}. Figure 1 gives the $\rho = \rho_{0}$ trajectories for the metric in Eq. (3) without the conformal factor, in Minkowski $\bar{R} - \bar{t}$ coordinates. They are the hyperbolae $\bar{R} = {\sqrt{\bar{t}^{2} + \rho_{0}^{2}}}$ with the null cones $\bar{R} = \pm{\bar{t}}$ as asymptotes and with acceleration $1/\rho_{0}$. The $t =$ const. curves are straight lines through the origin. The stretched horizon $\rho = \epsilon << 1/g$ corresponds to $\sqrt{-g_{tt}} = g \rho \rightarrow 0$. It is a timelike surface infinitesimally away from the event horizon $\rho = 0$. In other words, any $\rho = \rho_{0}$ hyperbola becomes a stretched horizon when $t \rightarrow \infty$ (it approaches the event horizon in this case \cite{TP2}).

 It is interesting to treat the region $\rho<b$ by means of the transformation
 \begin{equation}
 \rho' = \frac{b^{2}}{\rho}.
 \label{6}
 \end{equation}
 Inserting Eq. (6) in Eq. (3), we conclude that the metric in Eq. (3) is invariant under the inversion in Eq. (6). Therefore, we may say that the $\rho<b$ region is the image of the $\rho>b$ region obtained by using the inversion. This can be seen by constructing the conformal (Penrose) diagram corresponding to the geometry in Eq. (3). We use the fact that the causal structure of a spacetime is unchanged by conformal compactification \cite{PT}.
 
 Let us write the conformal metric in Eq. (3) in the form
 \begin{equation} 
  ds^{2} = g^{2} \rho^{2} \left( 1+\frac{b^{2}}{\rho^{2}}\right)^{2} \left[- dt^{2} + \left(\frac{d \rho}{g \rho}\right)^{2} + \frac{1}{g^{2}} \text{cosh}^{2} gt~ d\Omega^{2}\right].
 \label{7}
 \end{equation}
 The function $\Lambda^{-1}(\rho) = g^{2} \rho^{2} \left( 1+\frac{b^{2}}{\rho^{2}}\right)^{2}$ tends to zero when $\rho$ approaches infinity, so all points at infinity in the geometry of Eq. (7) can be brought at a finite distance in the new metric $d \bar{s}^{2} = \Lambda ds^{2}$ \cite{PT}. Discarding the conformal factor $\Lambda$, the new metric becomes 
 \begin{equation}
 d \bar{s}^{2} = -dt^{2} + d \bar{\rho}^{2} + \frac{1}{g^{2}} \text{cosh}^{2}gt ~d\Omega^{2},
 \label{8}
 \end{equation}
 where $d \bar{\rho} = d\rho/g \rho$; that is, $\bar{\rho} = (1/g) ln(\rho/b)$, where convenient initial conditions were used ($\bar{\rho} = 0$ at $\rho = b$). One observes that $\bar{\rho} \rightarrow \infty$ when $\rho >> b$ and that  $\bar{\rho} \rightarrow -\infty$ when $\rho << b$. We also note that the null geodesic $\rho = b e^{\pm{gt}}$ \cite{HC3} becomes $\bar{\rho} = \pm{gt}$ for the time - dependent Rindler metric in Eq. (8).
 
 We now compactify the spacetime in Eq. (8) by means of the new variables
 \begin{equation}
 u = \text{arctan}(t - \bar{\rho}),~~~~v = \text{arctan}(t + \bar{\rho}),
 \label{9}
 \end{equation}
 with finite ranges for the coordinates $u, v$, namely, $-\pi/2 \leq u,v \leq \pi/2$. The metric in Eq. (8) appears now as
 \begin{equation}
 d\bar{s}^{2} = -du dv + \frac{1}{g^{2}} \text{cosh}^{2} ~\frac{g(u+v)}{2} d\Omega^{2}.
 \label{10}
 \end{equation}
 The Penrose diagram is depicted in Fig. 2 (any point in the diagram is an expanding 2-sphere). The right half corresponds to $\rho > b$, and the left half to $\rho < b$. We observe that the diagram has no horizons.
 
 We know that the spacetime in Eq. (3) is not a solution of the vacuum Einstein's equations. Therefore, a source was introduced on the right-hand side \cite{HC3} in order for Eq. (3) to become an exact solution.

\section{ NONGEODESIC CONGRUENCE OF THE STATIC OBSERVER}

Let us consider now ''static'' observers in the spacetime of Eq. (3). Their 4 - velocity field is given by
\begin{equation}
u_{\alpha} = (-g \rho \omega, 0, 0, 0). 
\label{11}
\end{equation}
The scalar expansion $\Theta $ associated with $u_{\alpha}$ appears as
\begin{equation}
\Theta \equiv \nabla_{\alpha} u^{\alpha} = \frac{2~ \text{tanh}~ gt}{\omega \rho}.
\label{12}
\end{equation}
It is always finite and varies between $-2/\omega \rho$ and $2/\omega \rho$ when $t \rightarrow \mp \infty$.

The acceleration vector of the fluid world lines is given by
\begin {equation}
a^{\alpha} = (0, \frac{\rho^{2} - b^{2}}{\omega^{3} \rho^{3}}, 0, 0) ,
\label{13}
\end{equation}
with
\begin{equation}
\sqrt{a_{\alpha} a^{\alpha}} \equiv a(\rho) = \frac{|\rho^{2} - b^{2}|}{\omega^{2} \rho^{3}} .
\label{14}
\end{equation}
It is worth noting that $a^{\rho}$, from Eq. (13), changes its sign at $\rho = b$. This is in accordance with the interpretation of the surface $\rho = b$ as a domain wall and with the fact that $\Sigma(\rho)$ changes sign when the domain wall is crossed.

 One can associate to the hyperbolic ($\rho = \rho_{0})$ observer in the accelerated system a temperature $T(\rho_{0})$  given by $T(\rho_{0}) = a(\rho_{0})/2 \pi$ according to Unruh's formula. Even though the $\rho = 0$ hypersurface is no longer a horizon - compared to the $b = 0$ case (the geometry in Eq. (3) is time dependent and there is no a timelike Killing vector), we might formally construct a ''surface gravity'' as
\begin{equation}
\sqrt{a_{\alpha} a^{\alpha}} ~\sqrt{- g_{tt}}|_{\rho = 0} = g.
\label{15}
\end{equation}
Let us note that Abreu and Visser \cite{AV} defined a generalized surface gravity valid not only for a black - hole horizon but also for a sphere of arbitrary radius. The expression is also compatible with that for a time - dependent, spherically - symmetric situation, as in our case.\\
Our uniformly accelerating system is not static. However, for $\rho >> b$ or $\rho << b$, the ''proper'' temperature $T(\rho)$ obeys the Tolman relation \cite{MTW} 
\begin{equation}
T(\rho) \sqrt{-g_{tt}} = \frac{g}{2 \pi} \frac{|\rho^{2} - b^{2}|}{\rho^{2} + b^{2}} \approx \frac{g}{2 \pi} = const.,
\label{16}
\end{equation}
 where $T(\rho)$ is measured by a local observer. A temperature gradient is necessary to prevent heat to flow from regions with different gravitational potentials. We have, indeed, \cite{BH} 
 \begin{equation}
 Q_{\alpha} = - \kappa h_{\alpha}^{\nu} (T_{,\nu} + T a_{\nu}) = 0 ,
 \label{17}
 \end{equation}
 where $a_{\nu}$ is given by Eq. (13) , $\kappa$ is the coefficient of thermal conductivity and $Q_{\alpha}$ is the spacelike heat flux.
 
  The shear tensor of the flow,
  \begin{equation}
 \sigma_{\mu \nu} = \frac{1}{2} (h_{\nu}^{\alpha} \nabla_{\alpha} u_{\mu} + h_{\mu}^{\alpha} \nabla_{\alpha} u_{\nu}) - \frac{1}{3} \Theta h_{\mu \nu} + \frac{1}{2} (a_{\mu} u_{\nu} + a_{\nu} u_{\mu}), 
 \label{18}
 \end{equation}
has the nonzero components
\begin{equation}
\sigma_{\rho}^{ \rho} = -\frac{2 ~\text{tanh}~ gt}{3 \rho \omega} , ~~\sigma_{\theta}^{\theta} = \sigma_{\phi}^{\phi} = \frac{\text{tanh}~gt}{3 \rho \omega} ,
\label{19}
\end{equation}
 where $h_{\mu \nu} = g_{\mu \nu} + u_{\mu} u_{\nu} $ is the projection tensor in the direction perpendicular to $u_{\alpha}$. However, the rotation tensor $\omega_{\mu \nu}$ is vanishing. 

 The time - evolution of the scalar expansion is given by the Raychaudhuri equation
  \begin{equation}
 \frac{d \Theta}{d \lambda} - \nabla_{\alpha} a^{\alpha}+ \sigma^{\alpha \beta} \sigma_{\alpha \beta} - \omega^{\alpha \beta} \omega_{\alpha \beta}+ \frac{1}{3} \Theta^{2} = - R_{\alpha \beta} u^{\alpha} u^{\beta},
 \label{20}
 \end{equation} 
 where $\lambda$ is the parameter along the world lines. The Ricci tensor is defined by $R_{\alpha \beta} = g^{\mu \nu} R_{\alpha \mu \beta \nu}$, and 
 \begin{equation}
 \frac{d \Theta}{d \lambda} \equiv \dot{\Theta} = u^{\alpha} \nabla_{\alpha} \Theta = \frac{2}{\omega^{2} \rho^{2} \text{cosh}^{2} gt}.
 \label{21}
 \end{equation}
 Keeping in mind that
 \begin{equation}
 \sigma^{\alpha \beta} \sigma_{\alpha \beta} = \frac{2~\text{tanh}~gt}{3 \omega \rho},~~~\nabla_{\alpha} a^{\alpha} = \frac{2}{\omega^{4} \rho^{2}} (1 + \frac{b^{4}}{\rho^{4}}) ,
 \label{22}
 \end{equation}
 one obtains for the left-hand side of Eq. (20), the expression $-4b^{2}/(\omega^{4} \rho^{4})$, but $R_{tt}$ for the metric in Eq. (3) is given by $-4b^{2}/(\omega^{2} \rho^{2})$ \cite{HC3}. Using Eq. (11) we find that the Raychaudhuri equation is satisfied.

\section{BOUNDARY STRESS TENSORS}

 From the surface source term in the gravitational action, we know that the following energy - momentum tensor emerges  \cite{PW,EO} 
 \begin{equation}
 T_{\alpha \beta}^{(s)} = \frac{1}{8 \pi} (K_{\alpha \beta} - \gamma_{\alpha \beta} K),
 \label{23}
 \end{equation}
 where $(s)$ refers to the ''surface,'' $K_{\alpha \beta}$ is the extrinsic curvature tensor corresponding to $\rho = \rho_{0} =$ const. timelike hypersurface, $K$ is its trace and $\gamma_{\alpha \beta}$ is the induced metric given by 
 \begin{equation}
 \gamma_{\alpha \beta} = g_{\alpha \beta} - n_{\alpha} n_{\beta}.
 \label {24}
 \end{equation}
  The 4 - vector normal to the surface has the components $n^{\alpha} = (0, 1/\omega, 0, 0)$, where $~\omega = 1 + (b^{2}/\rho^{2})$.
 With the help of the well-known expression
 \begin{equation}
 K_{\alpha \beta} = \gamma_{\beta}^{\nu} \nabla_{\nu} n_{\alpha} ,
 \label{25}
 \end{equation}
 one obtains
 \begin{equation}
 K_{t}^{t} = K_{\theta}^{\theta} = K_{\phi}^{\phi} = \frac{\rho^{2} - b^{2}}{\omega^{2} \rho^{3}},
 \label{26}
 \end{equation}
 and the trace gives us
 \begin{equation}
 K \equiv \gamma^{\alpha \beta} K_{\alpha \beta} = \frac{3(\rho^{2} - b^{2})}{\omega^{2} \rho^{3}}.
 \label{27}
 \end{equation}
 Equation (23) now yields 
 \begin{equation}
 T_{t}^{(s)t} = T_{\theta}^{(s) \theta} = T_{\phi}^{(s) \phi} =- \frac{(\rho^{2} - b^{2})}{4 \pi \omega^{2} \rho^{3}}.
 \label{28}
 \end{equation}
 We see that the energy - momentum tensor is proportional to the metric tensor:
 \begin{equation}
 T_{\beta}^{(s) \alpha} = -\frac{(\rho^{2} - b^{2})}{4 \pi \omega^{2} \rho^{3}} \gamma_{\beta}^{\alpha},~~at~\rho = \rho_{0}. 
 \label{29}
 \end{equation}
 We have, therefore,
 \begin {equation}
 \Sigma(\rho) \equiv - T_{t}^{(s)t} = -p_{\theta} = -p_{\phi} = \frac{\rho^{2} - b^{2}}{4 \pi \omega^{2} \rho^{3}}, 
 \label{30}
 \end{equation}
 where $\Sigma(\rho)$ is the surface energy density and $p_{\theta}$ and $ p_{\phi}$  are the pressures (tensions for $\rho > b$)  on the angular directions. Note that, far from the Planck world ($\rho >> b),~\Sigma = 1/4 \pi \rho$. As a function of $\rho$, $\Sigma $ vanishes at $\rho = 0$ and $\rho = b$ (Fig. 3). In Cartesian coordinates, these correspond to the light cones $\bar{R} = \pm \bar{t}$ and the hyperbolic trajectories $\bar{R}~= \sqrt{\bar{t}^{2} + b^{2}}$, respectively.
 
 At $\rho_{01} = b (\sqrt{2} - 1)$, we have 
\begin{equation}
\Sigma_{min} = -\frac{2 - \sqrt{2}}{8 \pi b} < 0,
\label{31}
\end{equation}
and at $\rho_{02} = b (\sqrt{2} + 1)$, one obtains 
\begin{equation}
\Sigma_{max} = \frac{2 + \sqrt{2}}{8 \pi b} > 0 .
\label{32}
\end{equation}
$\Sigma $ is positive for $\rho>b$ and negative for $0<\rho<b$. The region around $\rho = b$ acts as a domain wall of thickness $\rho_{02} - \rho_{01} = 2b$. The fluctuations of the surface energy are completely negligible far from Planck's world. 

  Let us also observe that $a(\rho)$ from Eq. (14) obeys the relation 
\begin{equation}
a(\rho) = 4 \pi |\Sigma(\rho)| ,
\label{33}
\end{equation}
 in accordance with the Gauss law. A similar expression has been obtained in Ref. 22 for the Rindler horizon in Cartesian coordinates. For example, with $\rho_{0} = 1~ \text{cm} ~(\rho_{0} >> b)$, we obtain $\Sigma \approx 1/4 \pi \rho = 10^{48}~ \text{ergs/cm}^{2}$. One means our hyperbolic observer, located at $\rho_{0} = 1~ \text{cm}$ from the origin , perceives an enormous surface energy density. However, we must remember that the acceleration is also very large. On the contrary, we have $\Sigma = 10^{28}~ \text{ergs/cm}^{2} $ for $a = 10 ~\text{cm/s}^{2}$~ (with $~\rho_{0} \approx 10^{20}~ \text{cm}$).

 We find now the energy $E$ enclosed by the 2 - surface $H: \rho = \rho_{0}$ and $ t = t_{0}$. The metric tensor on H is
 \begin{equation}
 s_{\alpha \beta} = g_{\alpha \beta} - n_{\alpha} n_{\beta} + u_{\alpha} u_{\beta}. 
 \label{34}
 \end{equation}
 The energy is obtained from 
 \begin{equation}
 E = \int_{H} \Sigma(\rho) \sqrt{s} ~d\theta~ d\phi .
 \label{35}
 \end{equation}
 With $\sqrt{s} = \omega^{2} \rho^{2} \text{cosh}^{2} gt$ and $\Sigma(\rho)$ from Eq. (30), Eq. (35) yields
 \begin{equation}
 E = \rho (1 - \frac{b^{2}}{\rho^{2}}) \text{cosh}^{2} ~gt ,
 \label{36}
 \end{equation}
 taken at $\rho = \rho_{0}$ and $t = t_{0}$. It is worth noting here that $E$ becomes negative for $\rho < b$ and changes its sign under the inversion in Eq. (6).

 By means of the Padmanabhan formula $E = 2ST$ \cite{TP}, one obtains for the entropy $S$ corresponding to the degrees of freedom on $H$ (or adopting the holographic principle in the space enclosed by H) 
 \begin{equation}
 S = \frac{E}{2T} = \pi \omega^{2} \rho^{2} \text{\text{cosh}}^{2}~gt~~~~at ~\rho = \rho_{0} > b .
 \label{37}
 \end{equation}
 Hence, $S = A/4$, where $A$ is the area of $H$. This expression corresponds exactly to the horizon entropy of a black hole. That is not surprising since the $\rho =$ const. observer is at rest in the accelerated system, which is equivalent to a gravitational field. As Padmanabhan has noticed, one can construct a local Rindler observer who will perceive the timelike $\rho = \rho_{0}$ surface as a stretched horizon. The Unruh temperature on this surface will be $g/2 \pi \sqrt{-g_{tt}} = 1/2 \pi \rho_{0}$, the constant $g$ playing the role of the surface gravity. In addition, we can check that the expression in Eq. (36) is in accordance with the equipartition law \cite{TP} 
 \begin{equation}
 E = \frac{1}{2} n T ,
 \label{38}
 \end{equation}
 where $n$ is the number of elementary cells of area $l_{P}^{2}$ ( $l_{P}$ - the Planck length). Padmanabhan's formula applies here for $\rho > b$; otherwise, $E$ acquires a negative value. The case $\rho < b$ is not compatible with the statistical interpretation when the minimal area cell is $b^{2}$.

 The metric tensor $s_{\alpha \beta}$ of the  spacelike 2 - section $H$ of the constant $\rho$ hypersurfaces, to which the velocity vector $u^{\alpha}$ is normal, is given by (see Eq. (34))
 \begin{equation}
 s_{\alpha \beta} = diag(0, ~0, ~\omega^{2} \rho^{2} \text{cosh}^{2}gt, ~\omega^{2} \rho^{2} \text{cosh}^{2}gt~ \text{sin}^{2} \theta).
 \label{39}
 \end{equation}
 We may decompose the extrinsic curvature tensor $k^{\alpha}_{\beta} = s^{\nu}_{\beta} \nabla_{\nu} u^{\alpha}$ of $H$ into a traceless part and a trace part \cite{PW}: 
 \begin{equation}
 k_{\alpha \beta} = \sigma^{(2)}_{\alpha \beta} + \frac{1}{2} s_{\alpha \beta} \Theta^{(2)},
 \label{40}
 \end{equation}
 where $\sigma^{(2)}_{\alpha \beta}$ and $\Theta^{(2)}$ are, respectively, the shear tensor and the expansion of the surface elements. Using the components of the metric given by Eq. (39), we immediately find that $\sigma^{(2)}_{\alpha \beta}$ is vanishing and that $\Theta^{(2)}$ appears as
 \begin{equation}
 \Theta^{(2)} = s^{\nu}_{\alpha} \nabla_{\nu} u^{\alpha} = \frac{2~\text{tanh}~gt}{\omega \rho},
 \label{41}
 \end{equation}
 which is the same as Eq. (12). Therefore, the 2-surface stress tensor
 \begin{equation}
 t^{H}_{\alpha \beta} = \frac{1}{8 \pi} (k_{\alpha \beta} - k s_{\alpha \beta}) = - \frac{1}{16 \pi} \Theta^{(2)} s_{\alpha \beta}
 \label{42}
 \end{equation}
 has two nonzero components: 
 \begin{equation}
 t_{\theta}^{(H) \theta} = t_{\phi}^{(H) \phi} = - \frac{\text{tanh}~gt}{16 \pi \omega \rho}. 
 \label{43}
 \end{equation}
 
A comparison with a viscous Newtonian fluid leads to $\zeta = 1/16 \pi$, where $\zeta$ stands for the bulk viscosity coefficient. In other words, on the 2 - surface $H$, the pressure $(- \zeta \Theta^{(2)})$ is negative. Therefore, it acts as a surface tension. In spite of the fact that the authors of Refs. 1, 3, and 23 obtained a negative value for $\zeta$, our result is different because the 2 - surface H is not lightlike.

\section{CONCLUSIONS}

 We consider in this paper a uniformly - accelerated distribution of conformal Rindler observers and compute the stress tensors on 3 - and 2 - surfaces viewed as boundaries of spacetime. To any $\rho = \rho_{0}$ static observer in the accelerated system, we may associate a temperature proportional to its acceleration because a local Rindler observer will perceive the timelike surface as a stretched horizon.
 
 We have calculated the energy enclosed by a 2 - surface $H$ of constant $\rho$ and $t$ and have found that it obeys the equipartition law $E = (1/2) n T$. In addition, we have established that the entropy obtained from Padmanabhan's formula $S = E/2 T$ corresponds to the horizon of a black hole. That supports our view that $\rho = \rho_{0}$ observers are equivalent to those located near the black hole's horizon, where the geometry is of a Rindler type \cite{HC6}. The stress tensor on $H$ corresponds to a Newtonian viscous fluid with a bulk viscosity $\zeta = 1/16 \pi$, leading to a negative pressure that acts as a surface tension.

\begin{figure}
\includegraphics[width=15.0cm]{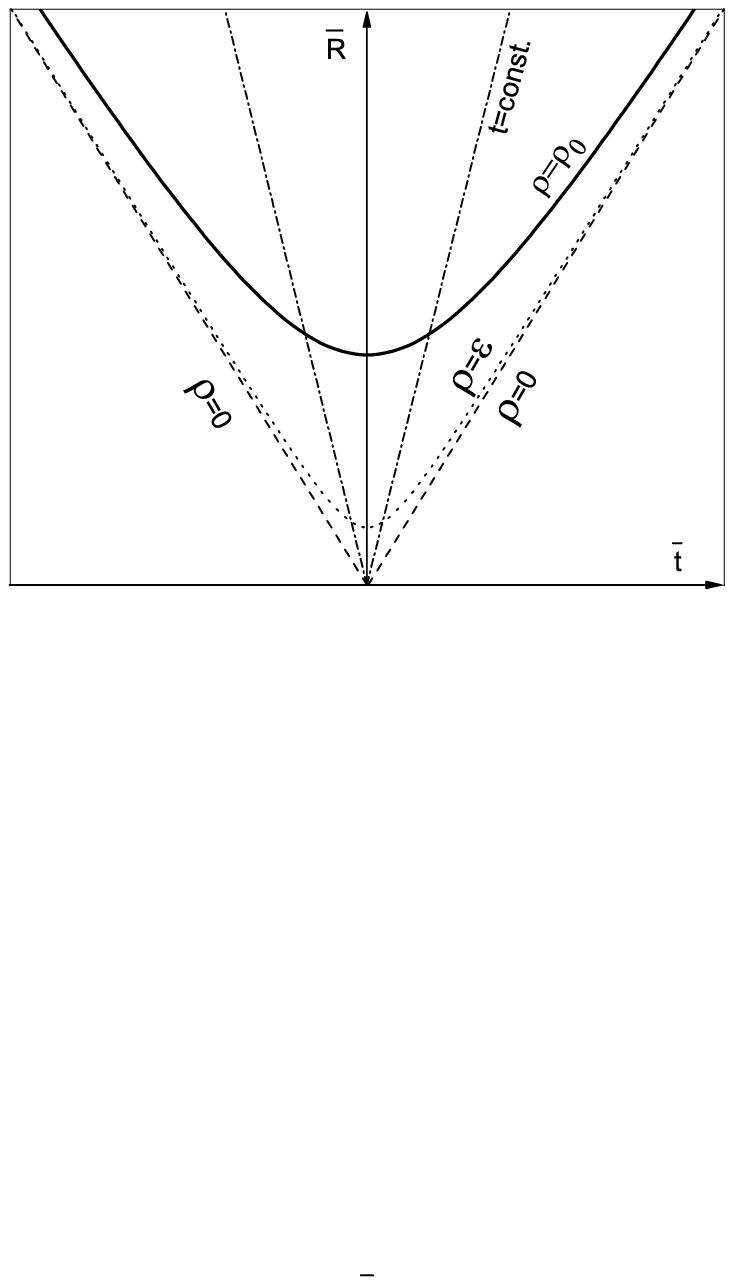}
\caption{  The $\rho = \rho_{0}$ (or $\bar{R} = {\sqrt{\bar{t}^{2} + \rho_{0}^{2}}}$) world line in Minkowski $\bar{R} - \bar{t}$ coordinates for the Rindler spacetime (the metric in Eq. (3) without the conformal factor). The solid curve is a hyperbola with $\bar{R} = \pm{\bar{t}}$ (dashed lines) as asymptotes. The $t$ = const. curves (dashed - dotted) are straight lines through the origin. $\rho = \epsilon << 1/g$ is the stretched horizon (dotted curve). The constant acceleration of the hyperbolic observer is $1/\rho_{0}$.}
\label{Fig.1}
\end{figure}

\begin{figure}
\includegraphics[width=15.0cm]{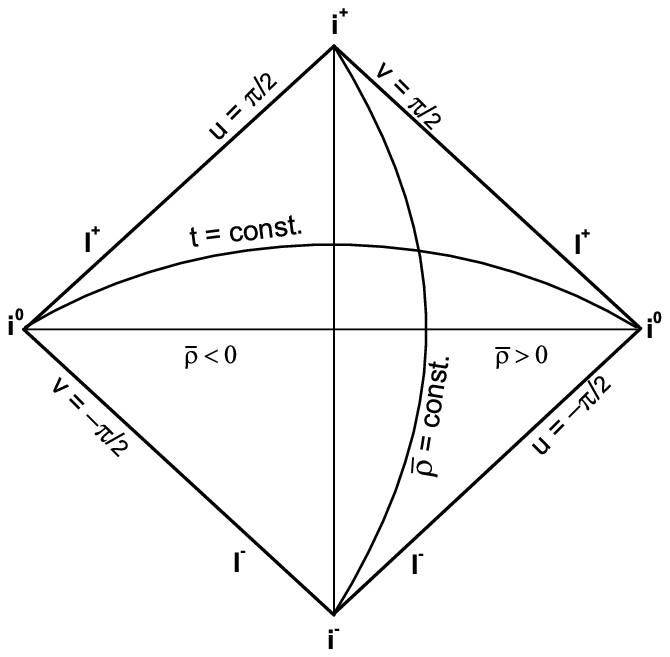}
\caption{ Conformal diagram for spacetime (7). The compact coordinates $u$ and $v$ run from $- \pi/2 $ to $\pi/2$. The diagram covers both the $\rho < b~(\bar{\rho} < 0)$ region and the $\rho > b~(\bar{\rho} > 0)$ region. The $u = \pm{\pi/2}$ and $v = \pm{\pi/2}$ solid null lines are shown, together with the timelike $\bar{\rho} = const.$ and the spacelike $t = const.$ curves. $i^{-}$ and $i^{+}$ are, respectively, the past timelike infinity and the future timelike infinity, $i^{0}$ is the spacelike infinity and $I^{-}$ and $I^{+}$ are the past null infinity and the future null infinity, respectively. Every point on the figure is an expanding $S^{2}$.}
\label{Fig.2}
\end{figure}

\begin{figure}
\includegraphics[width=15.0cm]{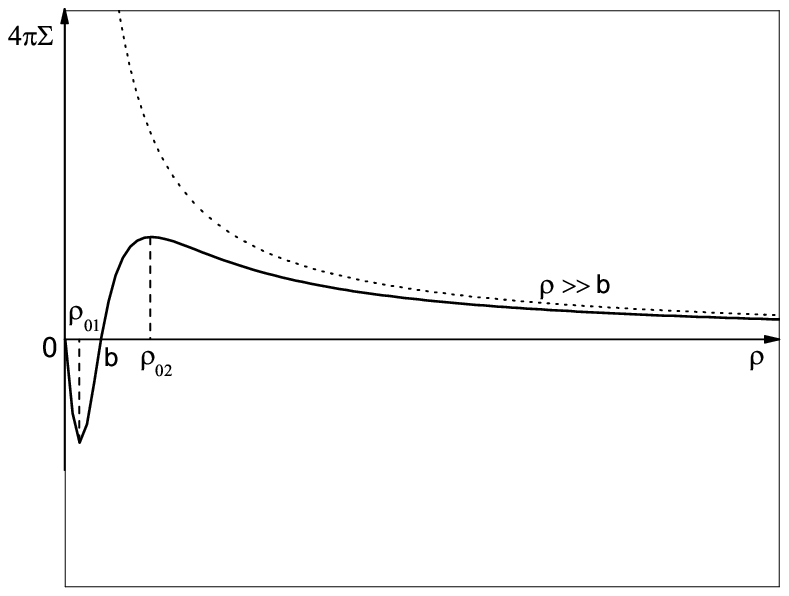}
\caption{ Surface energy density $\Sigma$ versus the radial coordinate $\rho$. $\Sigma$ becomes negative for $\rho < b$ and has two extrema, one at $\rho = \rho_{01} = b(\sqrt{2} -1)$ and the other at $\rho = \rho_{02} = b(\sqrt{2} + 1)$. The dotted curve represents the hyperbola $\Sigma(\rho) = 1/2 \pi \rho$ ($\rho >> b$).}
\label{Fig.3}
\end{figure}


\begin{thebibliography} {24}
\bibitem {KP}
J. Khoury and M. Parikh, ArXiv: hep-th/0612117.
\bibitem{TPM}
K. S. Thorne, W. H. Zurek and R. H. Price, \textit{Black Holes: The Membrane Paradigm}, edited by K. S. Thorne, R. H. Price and D. A. Macdonald (Yale University Press, New Haven, CT, 1986); M. S. Morris and K. S. Thorne, Am. J. Phys. \textbf{56}, 395 (1988).
\bibitem{PW}
M. Parikh and F. Wilczek, Phys. Rev. D\textbf{58}, 064011 (1998).
\bibitem{HC1}
H. Culetu, Cent. Eur. J. Phys. \textbf{6}, 317 (2008).
\bibitem{OA}
O. Aharony, S. S. Gubser, J. Maldacena, H. Ooguri and Y. Oz, Phys. Rep. \textbf{323}, 183 (2000) (ArXiv : hep-th/9905111); M. Fujita, JHEP \textbf {0810}, 031 (2008) ( ArXiv : 0712.2289 [hep-th]).
\bibitem{BY}
J. D. Brown and J. W. York, Jr., Phys. Rev. D\textbf{47}, 1407 (1993).
\bibitem{HC2}
H. Culetu, Proceeding \textit{AIP Conference Proceedings} (The British University, Cairo, 2009), p. 129; ArXiv: 0804.3754 [hep-th].
\bibitem{HC3}
H. Culetu, ArXiv: 0905.3474 [hep-th].
\bibitem{TP}
T. Padmanabhan, ArXiv: 0912.3165 [gr-qc].
\bibitem{EV}
E. Verlinde, ArXiv: 1001.0785 [hep-th]; F. Caravelli and L. Modesto, ArXiv: 1001.4364 [gr-qc]; H. Culetu, ArXiv: 1002.3876 [hep-th]; J-W. Lee, H. Kim and J. Lee, ArXiv: 1002.4568 [hep-th]; J-W. Lee, ArXiv: 1003.4464 [hep-th]: Y. S. Myung, ArXiv: 1003.5037 [hep-th]; Y-F. Cai, J. Liu and H. Li, 1003.4526 [astro-ph].
\bibitem{EO}
C. Eling and Y. Oz, ArXiv: 0906.4999 [hep-th].
\bibitem{EW}
E. Witten, Nucl. Phys. B\textbf{195}, 481 (1982).
\bibitem{HC4}
H. Culetu, Gen. Relativ. Gravitation \textbf{26}, 282 (1994).
\bibitem{CP}
Y. M. Cho and D. G. Pak, ArXiv: 1001.4925 [hep-th].
\bibitem{SWH}
S. W. Hawking, Phys. Lett. B\textbf{195}, 337 (1987).
\bibitem{SW}
S. Weinberg, Rev. Mod. Phys. \textbf{63}, 1 (1989).
\bibitem{TP2}
T. Padmanabhan, Int. J. Mod. Phys. D\textbf{18}, 2189 (2009).
\bibitem{PT}
P. Townsend, ArXiv: gr-qc/9707012; C. Semay, Eur. J. Phys. \textbf{28}, 877 (2007).
\bibitem{AV}
G. Abreu and M. Visser, ArXiv: 1005.1132 [gr-qc]; ArXiv: 1004.1456 [gr-qc].
\bibitem{MTW}
C. Misner, K. Thorne and J. A. Wheeler, \textit{Gravitation} (Freeman, San Francisco, 1973).
\bibitem{BH}
I. Brevik and K. T. Heen, Astrophys. Space Sci. \textbf{219}, 99 (1994).
\bibitem{HC5}
H. Culetu, Int. J. Mod. Phys. D\textbf{15}, 2177 (2006).
\bibitem{TD}
T. Damour, \textit{Proceedings of the 2nd Marcel Grossmann Meeting on General Relativity} (North Holland Publishing Company, Shanghai, 1982), p. 587.
\bibitem{HC6}
H. Culetu, ArXiv: 0903.3548 [hep-th].


\end{thebibliography}
\end{document}